\documentclass[preprint,aps,amsmath,showkeys,showpacs,floatfix]{revtex4}
\usepackage{graphicx}
\usepackage{dcolumn}
\usepackage{bm}

\begin{document}


\title{More about the Laughlin droplet}
\author{N. Barber\'an}
\affiliation{Departament d'Estructura i Constituents de la Mat\`eria,
Facultat de F\'\i sica, Universitat de Barcelona,
E-08028 Barcelona, Spain}

\vskip4mm
\begin{abstract}
We compare the exact diagonalization ground state wave function
(calculated without
any restriction) of a two-dimensional droplet in a
perpendicular magnetic field with the Laughlin ansatz,
as the number of electrons increases. The
fully spin-polarized
case for filling factor $1/3$ without lateral confining potential or
Zeeman effect is considered. We observe that the overlap decreases as
the number of electrons increases.
\end{abstract}
\vskip2mm

\maketitle
\vfill
\eject


Most of the knowledge and understanding of the fractional quantum
Hall effect states is due to the existence of  Laughlin's wave
function \cite{lau}, a simple and intuitive analytical expression that
matches very well with the ground states (gs) of the system
characterized by the filling factor $\nu=1/q$, where $q$ is an odd
integer. This has been tested by numerical calculations of
finite systems for different geometries \cite{hal,hal2,he,rez}
as well as experimental results related with measurable quantities
which can be derived from its properties, as is the case of the
fractional charge of some of the excited states \cite{gol}.
We analyze the overlap between this ansatz and the gs
wave function that comes out from numerical
exact diagonalization of the Hamiltonian for finite systems. We also
consider other properties of the gs.

\medskip

Laughlin's wave function for $\nu=1/q$
is given by:
\begin{equation}
\Psi_L=\prod_{i<j} (z_i-z_j)^q \,\,e^{-\sum \mid z_i\mid^2/4}
\end{equation}
where $z=x+iy$ (in units of $l_B=\sqrt{\frac{\hbar c}{eB}}$, $B$
being the magnetic field) and where the symmetric gauge has been
assumed. The
physical meaning is easily deduced from its form: the minimization of
the energy in the partially filled lowest Landau level is obtained
when each electron sees zeros of order
$q$ at
the positions of the other electrons in such a way that there are non
free zeros which would increase the energy. It can be verified that if
the polynomial part is expanded, one obtains a combination of Slater
determinants built up from Fock Darwin \cite{jac} single particle
wave functions of the type:
\begin{equation}
\phi_{sp}\sim z^m \,\,e^{-\mid z\mid^2/4}
\end{equation}
(solutions of the non-interacting system) where $m$ is the single
particle angular momentum. The expansion is a
homogeneous polynomial of degree $M$ (the total angular momentum) and
contains single particle angular momenta up to a maximum given
by,
\begin{equation}
m_L=q(N-1)
\end{equation}
where $N$ is the number of electrons.
In order to analyze the evolution of the wave function given by Eq.(1)
with
$N$, we consider the cases
$N=2$, $3$ and $4$ keeping the condition given by
$\nu=1/3$ . We compare Eq.(1) with
the wave function that comes out from the exact diagonalization (we
only considered the fully spin polarized case). In the exact
calculation no Zeeman
or kinetic energy contributions are considered, electrons are confined
within a disk due to the restriction on the total angular
momentum.

\medskip

For $N=2$ the dimension of the subspace characterized by the well
defined quantum numbers $M=3$, the angular momentum needed to
obtain $\nu=1/3$ for $N=2$
($M=qN(N-1)/2$)
and total spin along the z-direction
$S_z=1$ is equal to two, that is to say, the exact
gs is a combination of two Slater determinants in which the largest
possible single particle angular momentum is given by,
\begin{equation}
m_{ex}=M-\frac{(N-1)(N-2)}{2}
\end{equation}
or $m_{ex}=3$ in this case, in agreement with the value obtained from
Eq.(3). Furthermore, the linear combination of Slaters is the same in
both cases (in Eq.(1) and in the diagonalization) and consequently the
overlap is equal to one. That is to
say, for $N=2$ the Laughlin wave functiuon is exact. For $N=3$ ($M=9$)
the situation is not the same: the number of Salters involved in the
g.s. is not the same, the maximum $m$ is also different and as a
consequence the overlap is lower than one. This tendency is increased
for $N=4$. A summary of the results is given in Table 1 and Figs.1 and
2 below. $N_L$ and $N_{ex}$ are the total number of Slater
determinants included in the Laughlin and exact ground states
respectively, i.e., $N_{ex}$ is the dimension of the Hilbert space
characterizad by $M$ and $S_z$. $P_L$ and
$P_{ex}$ are the normalized weights of the most important
Slater within the expansion of the g.s. respectively. This Slater
determinant contains, for all
the values of $N$ considered, a packet of successive single particle
angular momenta separated from the center of the dot in such a way
that they produce a compact ring. Remarkably this compact
structure made up of successive single particle angular
momenta wavefunctions ( $m=2,3$ and $4$ for $N=3$) differs from
the structure intuitively suggested by that of electrons
surrounded by magnetic quantum fluxes. Namely, the zeros at the
electronic positions seem to emphasize the short range character
of the interaction rather than the separation between electrons. Fig.1
shows the overlap
$S=\mid
\langle\Psi_L\mid
\Psi_{ex}\rangle\mid^2$ for $N=2$, $3$ and $4$ and Fig.2 shows the
values of
$P_L$ and $P_{ex}$ for different N.
The
tendency seems to indicate that the overlap will decrease as $N$
increases and so, as the function given by Eq.(1)
is a
especially good approximation for a low number of electrons, it must
be taken with some care for large $N$ in a finite system.

\medskip

It must be emphasized that the tendency of the overlap to worsen as
the number of electrons increases would be a trivial result if the
gs were a single Slater determinant and also if we used
differet single particle wavefunctions: the exact wavefunctions for
$\Psi_{ex}$ and approximate wavefunctions for $\Psi_L$. In this case,
the overlap would scale as $(1-\epsilon)^N$ $\epsilon$ being a small
number for good trial functions. However this is not our case: the
gs is a linear combination of several Slater determinants
and in addition, we use the same Slaters to build up the linear
combinations in each case, that is to say, the overlap depends on the
coefficients, namely, on the electron-electron interaction and is not
directly related to $N$. As a consequence, for example, Laughlin's
ansatz is the exact result for filling factor $\nu=1$ independently of
$N$, since only one determinant is involved.

\medskip

Special attention must be paid to the increasing difference between
$N_L$ and $N_{ex}$ due to the fact that the ratio between the
weights of
the Slater determinants that are lacking in $\Psi_L$ related to the
weights of the Slaters which are included is not negliglible.
As an example, for $N=4$ one of the Slaters lacking in the $\Psi_L$
has a normalized weight of $0.002$ compared with that of the
most
important one which is $0.331$. Furthermore, $\Psi_L$ loses the
small
and the large values of $m$ and so the center as well as the edge of
the droplet are poorly reproduced. As an example for $N=5$, there are
$27$ Slaters that contain $m=0$ and only $16$ of them are included in
the expansion of $\Psi_L$. However, in spite of the fact that the
overlap of
the gs wavefunction and Laughlin's ansatz worsens as $N$
increases, it is not the case for some expected values, especially
for the energy. If $H$ has only the Coulomb contribution  (as is
appropriate for a fully spin polarized system in the lowest
Landau level regime), and if we define the discrepancy $D$ as
$[(E_{ex}-E_L)/E_{ex}]\,\times\,100$ where
$E_{ex}=\langle\Psi_{ex}\mid H\mid \Psi_{ex}\rangle$ and $E_L=\langle
\Psi_L\mid H \mid \Psi_L\rangle$ the results obtained are:

\medskip

For $N=2$, $D=0$, $D=10.5$ for $N=3$ and $D=1.3$ for $N=4$, i.e.
improving the result as $N$ increases.

\medskip

There are several previous studies on the Laughlin wave function for
finite systems which devoted special attention to the study of the
edge states: Mitra and MacDonald
\cite{mit} have analyzed the angular
momentum distribution function for a droplet and found that the
occupations are peaked at the edge (for $N=15,20$ and $25$) and
that it has a rapid decline. We believe that it can be a consequence
of
the reduced base implied in $\Psi_L$ as it loses the small and large
values of the single particle angular momenta. Tsiper and Goldman
\cite{tsi} compare the density of a droplet obtained from Eq.(1) and
from exact diagonalization for $N=5$ to $12$ and $\nu=1/3$. They
obtain important differences at the centre of the droplet and a
nearly
exact coincidence at the edge. The difference in the electron-electron
interaction
implied in the Laughlin ansatz (short range interaction) and in the
exact calculation (Coulomb interaction) is invoked in order to
explain the difference and the formation of striped states. We believe
that their result could be a consequence of the procedure used to
obtain the results: within the exact calculation they truncate the
base of the Hilbert space, i.e.,
the $m^{'}_{ex}$ considered is obtained by increasing Hilbert
space until overlap $S$ converges to at least three significant
digits. However this procedure forces a precise coincidence with
$\Psi_L$ at the edge, giving no information about the total weight
of the rest
of the members of the base of the Hilbert space, which is not
necessarily negligible, as we mentioned previously. As an example, for
$N=12$ they consider $m^{'}_{ex}=35$ ($m_L=33$) while $m_{ex}=143$
is the exact single-perticle maximum angular momentum involved. At the
center, their numerical calculation contains all the Slaters without
restriction and they obtain strip-like oscillations on the radial
electron densities which has been
proved to be responsible for the observed unexpected behavior of the
current-voltage power law \cite{yan,wan}.

\medskip

Finally we conclude that for finite systems the overlap between the
Laughlin wave function and the exact results is worse as $N$
increases. This result is similar to that obtained previously by
Haldane \cite{hal} for spherical geometry that mimics a two
dimensional
homogeneous system. His conclusion was that for $N=3$ the Laughlin
type function is the exact solution ( even for Coulomb interaction)
but it is not for $N\geq 4$. In a recent paper by Yannouleas and
Landman \cite{yan1} a systematic study of a system of $6$
electrons in a range of filling factors from $\nu=1/5$ to
$\nu=1/9$ is reported. They conclude that the analytical model of
collectively rotating electron molecules (REM) \cite{yan} provides
better representation of the system. Other references
\cite{hal,he} have
tested some results obtained with Eq.(1) for some particular values
of $N$, however our aim is to study
the tendency
as $N$ increases. The contribution of our report refers
to the edge as well as the central properties of finite systems which
can be appreciably different from those properties
obtained by the use of the Laughlin droplet for large $N$ and as a
consequence, the differences can be significant at the thermodynamic
limit. However, the evolution of the overlap of the wave functions
does not necessarily characterize the evolution of the expected values
of
some operators as was previously pointed out for the eneregy operator.
\medskip

\vskip6mm

We gratefully acknowledge F. Salvat and J. Soto for helpful
discussions. This work has been performed under Grants No.
BFM2002-01868 from DGESIC (Spain) and No. 2001GR-0064 from Generalitat
de Catalunya.

\eject

\eject

\begin{figure}[htb]
\includegraphics*[width=0.7\columnwidth]{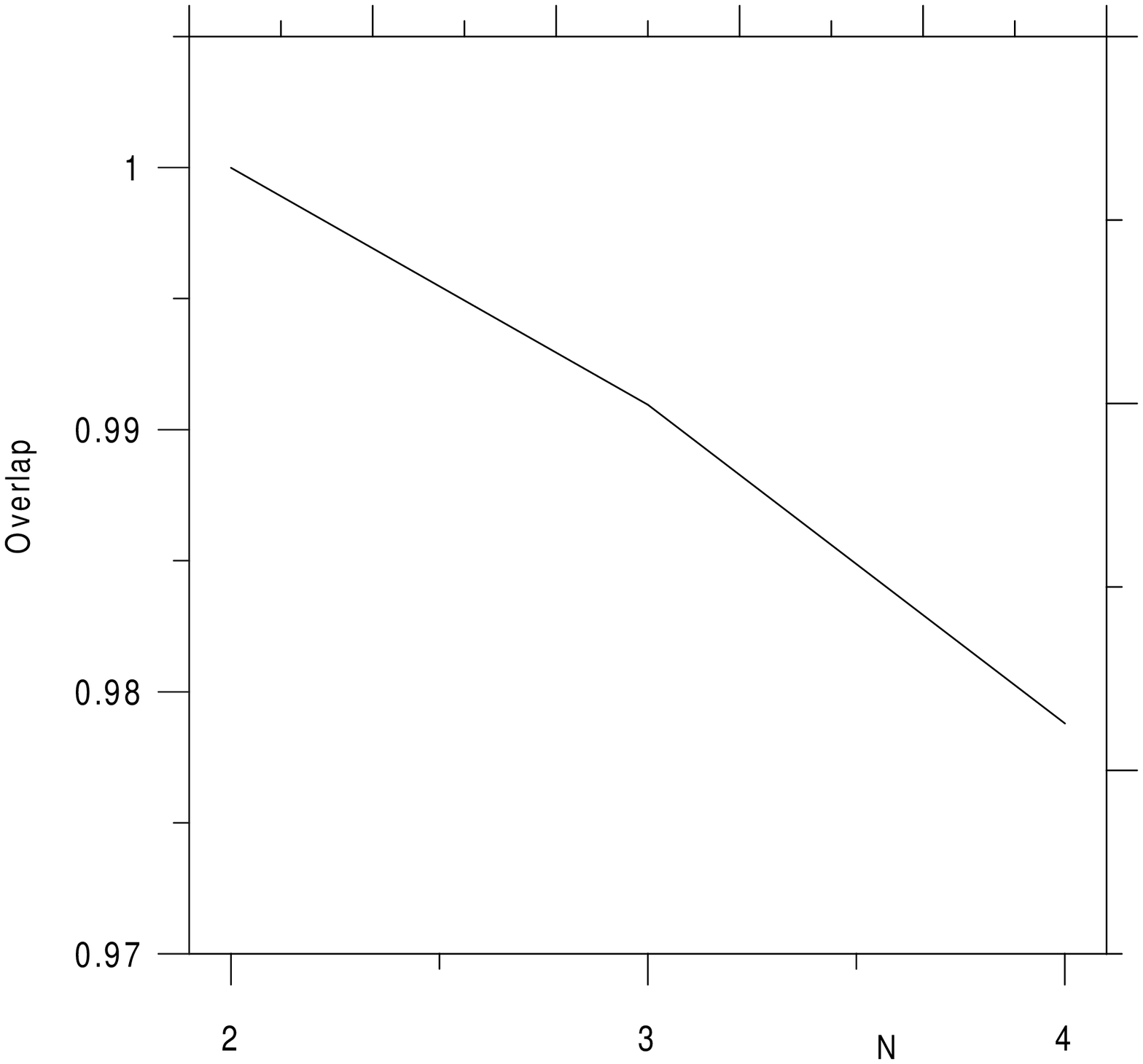}
\caption{Overlap between the Laughlin wave function and the exact
diagonalization result as a function of $N$}
\end{figure}
\eject

\begin{figure}[htb]
\includegraphics*[width=0.7\columnwidth]{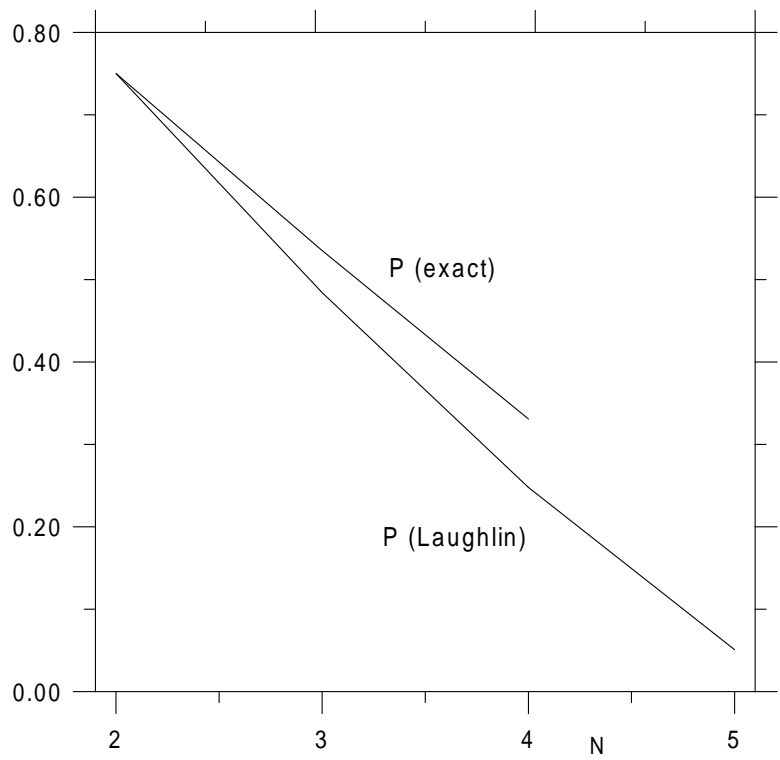}
\caption{Evolution of the normalized weight of the most important
Slater determinant within the wave function expansion as $N$
increases}
\end{figure}
\eject

\begin{figure}[htb]
\includegraphics*[width=0.7\columnwidth]{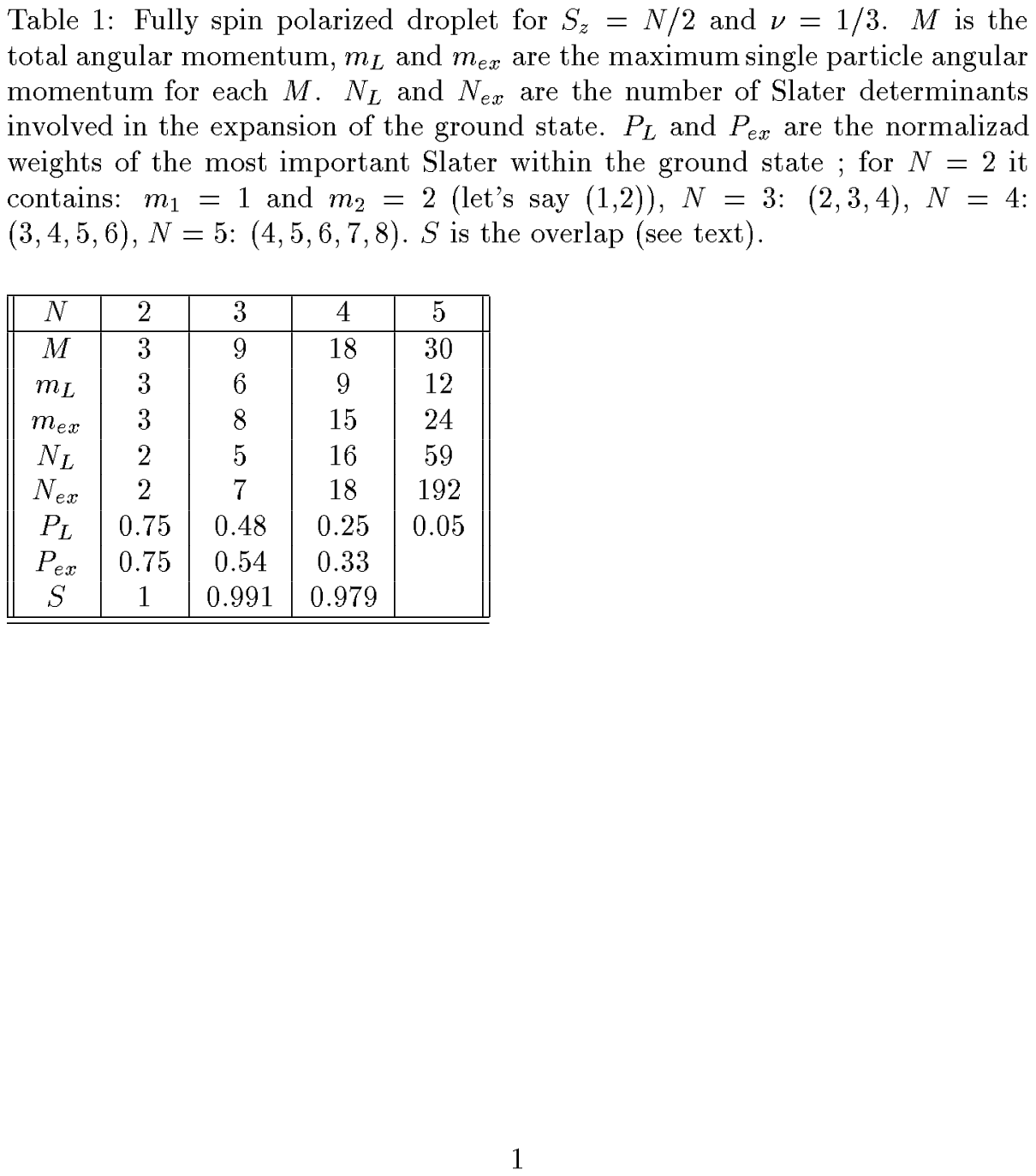}
\end{figure}

\end{document}